\documentclass[preprint,amsmath,amssymb,prl]{revtex4}
\usepackage{graphicx}% Include figure files
\usepackage{dcolumn}% Align table columns on decimal point
\usepackage{bm}% bold math
\newcommand{\pd}{\partial}

\begin{document}

\preprint{To be published in Nature on November 16, 2006}

\title{Hydrodynamic turbulence cannot transport angular momentum effectively in astrophysical disks}

\author{Hantao Ji$^1$}
%\email{hji@pppl.gov}
\author{Michael J. Burin$^{1,\dagger}$}
\author{Ethan Schartman$^1$}
\author{Jeremy Goodman$^1$}
\affiliation{$^1$Center for Magnetic Self-organization in Laboratory and Astrophysical Plasmas,
Plasma Physics Laboratory and
Department of Astrophysical Sciences, Princeton University, Princeton, New Jersey 08543, USA\\
$^\dagger$Present address: Department of Physics and Astronomy, Pomona College, Claremont, California 91711, USA}

%\date{Received 8 August; accepted 6 October 2006}

\maketitle

The most efficient energy sources known in the Universe are accretion disks. Those around black holes convert 5 -- 40 per cent of rest-mass energy to radiation. Like water circling a drain, inflowing mass must lose angular momentum, presumably by vigorous turbulence in disks, which are essentially inviscid \citep{shakura73}.
The origin of the turbulence is unclear. Hot disks of electrically conducting plasma can become turbulent by way of the linear magnetorotational instability \citep{balbus98}.  Cool disks, such as the planet-forming disks of protostars, may be too poorly ionized for the magnetorotational instability to occur, hence essentially unmagnetized and linearly stable.
Nonlinear hydrodynamic instability often occurs in linearly stable flows (for example, pipe flows) at sufficiently large Reynolds numbers. Although planet-forming disks have extreme Reynolds numbers, Keplerian rotation enhances their linear hydrodynamic stability, so the question of whether they can be turbulent and thereby transport angular momentum effectively is controversial
\citep{zeldovich81,dubrulle93,balbus96,richard99,richard01,longaretti02,
chagelishvili03,yecko04,umurhan04,garaud05,mukhopadhyay05,dubrulle05b,lesur_longaretti05}.
Here we report a laboratory experiment, demonstrating that non-magnetic quasi-Keplerian flows at Reynolds numbers up to millions are essentially steady. Scaled to accretion disks, rates of angular momentum transport lie far below astrophysical requirements. By ruling out purely hydrodynamic turbulence, our results indirectly support the magnetorotational instability as the likely cause of turbulence, even in cool disks.

Our experiments involved a novel Taylor-Couette apparatus\citep{taylor23}.  
The rotating liquid (water or a water/glycerol mixture) is confined between two
concentric cylinders of radii $r_1,r_2$ ($r_2>r_1$) and height $h$.  The
angular velocity of the fluid is controlled by that of
the cylinders, $\Omega_1$ and $\Omega_2$.
An infinitely long, steady, Taylor-Couette flow rotates as
\begin{equation}
\Omega(r)=a+\frac{b}{r^2},
\label{Couette}
\end{equation}
where $a = (\Omega_2 r_2^2 - \Omega_1 r_1^2)/(r_2^2-r_1^2) $ and $b =
r_1^2r_2^2(\Omega_1-\Omega_2)/(r_2^2-r_1^2).$ 
Reynolds number here can be defined as
$\bar r (r_2-r_1)(\Omega_1-\Omega_2)/\nu$, where $\nu$ is viscosity and 
$\bar r\equiv(r_1+r_2)/2$. 
The rotation profile
(equation (\ref{Couette})) ensures a radially constant viscous torque
$-2\pi\rho\nu h r^3\partial\Omega/\partial r$ for constant mass density
$\rho$ and constant $\nu$.  Astrophysical disks are mostly
Keplerian, meaning $\Omega\propto r^{-3/2}$, so that $|\Omega|$
decreases radially outward ($\pd |\Omega|/\pd r<0$) while the specific angular momentum, 
$|r^2\Omega|$, increases radially ($\pd |r^2\Omega|/\pd r>0$).
We apply the term ``quasi-Keplerian'' to
any flow satisfying these conditions, which are crucial for
both hydrodynamic and magnetohydrodynamic linear stability
\citep{balbus98}.

\begin{figure}[]
\centerline{
\includegraphics[width=2.5truein]{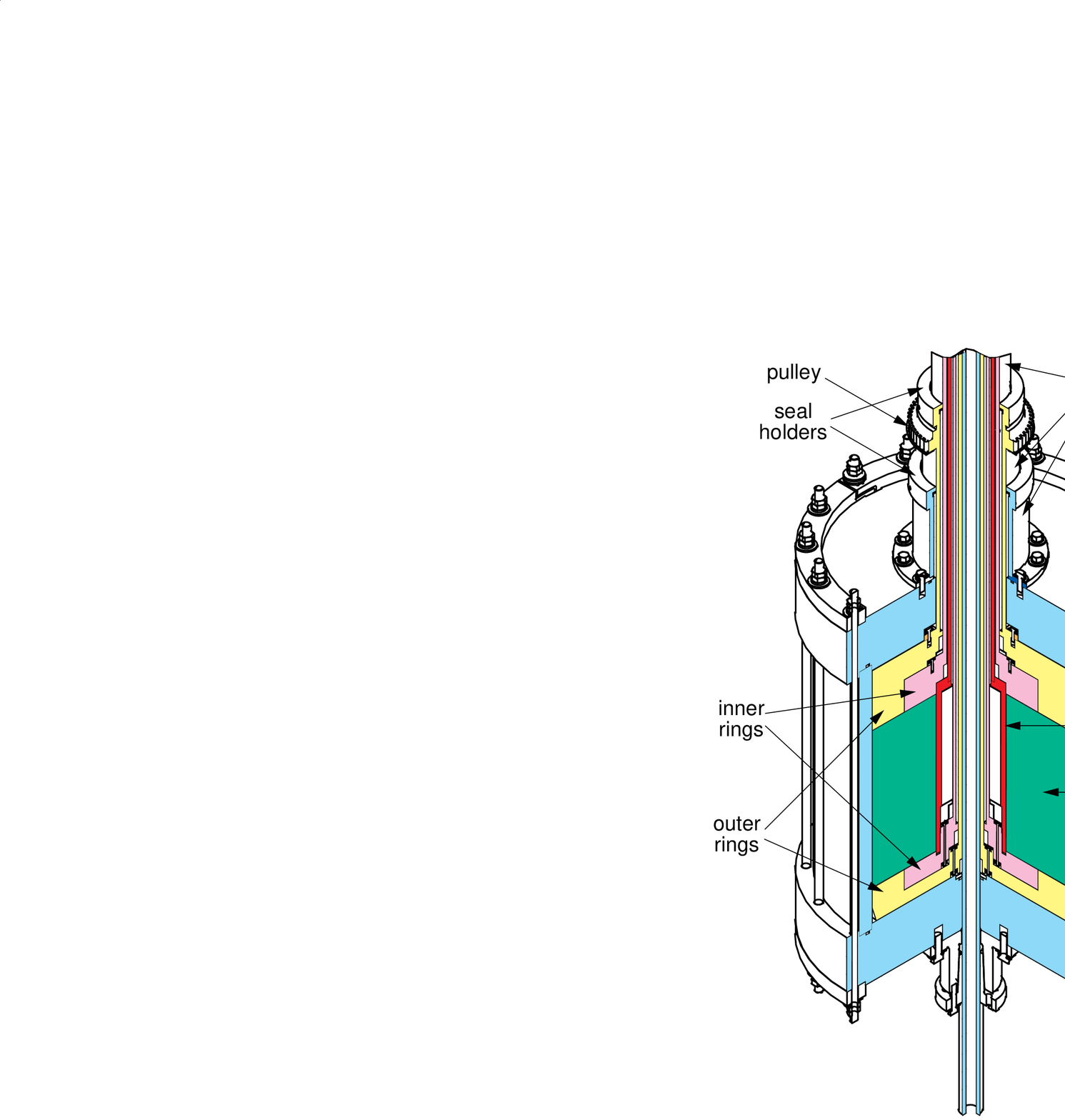}}
\caption{\textbf{Experimental setup.}
A rotating fluid (water or a water/glycerol mixture) of height $h=27.86$cm is confined between two
concentric cylinders of radii $r_1=7.06$cm and $r_2=20.30$cm,
which rotate at rates of $\Omega_1$ and $\Omega_2$, respectively.
Two novel features distinguish this apparatus from
conventional Taylor-Couette experiments.
First, secondary circulation is controlled
by dividing each endcap into two independently driven rings.
Opposing rings at top and bottom are driven at the same selectable angular velocity
$\Omega_3$ (inner rings) and $\Omega_4$ (outer rings).
Traditionally, a large aspect ratio $\Gamma \equiv h/(r_2-r_1)$ is used
to reduce the secondary circulation. 
However, at $\Gamma \simeq 25$ by Richard \citep{richard01} and even at $\Gamma \simeq
100$ by Taylor \citep{taylor36}, end effects were reported to be significant when
the endcaps co-rotated with one of the cylinders. Even when the endcaps were divided
into two rings, but with each affixed to one cylinder \citep{wendt33,richard01},  
residual secondary circulation may have facilitated observed turbulent
transitions\citep{schultz59,lesur_longaretti05}. 
When $\Omega_3$ and $\Omega_4$ are appropriately chosen,
secondary circulation is minimized and ideal Couette
profiles are well approximated.\citep{burin06}.
A second novel feature is access to rotation profiles on both sides of marginal
linear stability at Reynolds numbers as large as $10^6$ [also see Fig.(\ref{figure:diagram})]. 
When the specific angular momentum, $r^2\Omega$, 
decreases with increasing $r$, the Rayleigh stability criterion\citep{rayleigh16} is
violated, and thus the flow is linearly unstable when Reynolds number 
exceeds a critical value\citep{taylor23}.
When $\pd|r^2\Omega|/\pd r>0$ but $\pd|\Omega|/\pd r<0$ 
(as in disks, where $\Omega\propto r^{-3/2}$),
the flow is quasi-Keplerian and known to be linearly axisymmetrically stable. 
All major components of the apparatus were precisely machined and balanced, 
and except for the inner cylinder and rotating shafts,
are made of clear acrylic to facilitate visual and laser diagnostics.} 
\label{figure:setup}
\end{figure}

Real flows have finite length. Disks have nearly stress-free vertical boundaries, but viscous stress at the vertical endcaps of laboratory flows drives secondary circulation. This may cause the rotation profile to deviate significantly from equation (\ref{Couette}), and may even provoke turbulence
\citep{wendt33,taylor36,schultz59,richard01,
kageyama04,lesur_longaretti05}.
Our apparatus incorporates two rings at each end that are driven
independently of the cylinders (Fig.\ref{figure:setup}).  Thus we have
four controllable angular velocities.  When we choose these
appropriately, secondary circulation is minimized, and ideal Couette
profiles are closely approximated throughout the flow except within
$\sim 1$~cm of the endcaps\citep{burin06}.

Most past work has taken the outer cylinder at rest ($\Omega_2=0$), so
that both $|\Omega|$ and $|r^2\Omega|$
decrease radially outward.  Such flows are axisymmetrically linearly
unstable\citep{taylor23} at modest Reynolds number, $Re$. Very few
experiments have studied the linearly stable regime where
$\pd|r^2\Omega|/\pd r>0$, as occurs in disks.  Among these few are two
classic experiments of the 1930's: in one of these, the inner cylinder
was at rest ($\Omega_1=0$)\citep{taylor36}, while in the other
\citep{wendt33}, $0 \leq \Omega_1/\Omega_2 \leq 1$.  Enhanced torques
between the cylinders and other evidence of turbulence were reported
at sufficiently large $Re$.  These results have been cited in support
of the hypothesis of nonlinear hydrodynamic turbulent transport in disks
\citep{zeldovich81,richard99}, notwithstanding that the
direction of turbulent angular momentum transport,
which always follows $-\pd\Omega/\pd r$ on energetic grounds, differs in
sign between these experiments and astrophysical disks.
The so-called $\beta$ prescription\citep{richard99} derived
from the above experiments has been used to model or interpret
astronomical observations\citep[e.g.][]{hueso05}.  To our knowledge,
the only published experiments with quasi-Keplerian flow at relevant
Reynolds numbers are those of Richard\citep{richard01,dubrulle05b} and
Beckley\citep{beckley02}.  In Richard's work, transition
to turbulence via nonlinear instabilities was studied qualitatively by
a flow-visualization method. No direct measurements of angular
momentum transport were performed.  Beckley did measure torques
roughly consistent with the $\beta$ prescription
but attributed them to secondary circulation, which was strong
because his endcaps co-rotated with the outer cylinder and $h/(r_2-r_1)$ was
only $\sim 2$.

\begin{figure}[]
\centerline{
\includegraphics[width=3.6truein]{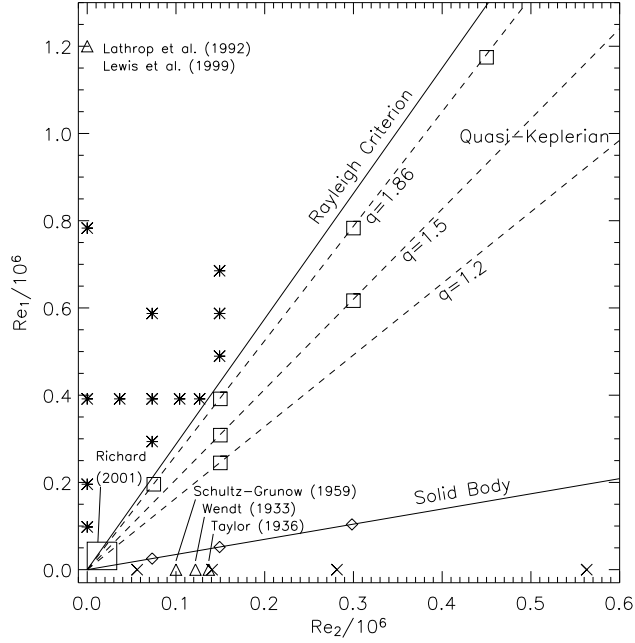}}
\caption{\textbf{Experimentally studied Taylor-Couette flows.}  Axes are Reynolds numbers based
  on the inner and outer cylinders: $Re_{1,2} \equiv \Omega_{1,2}
  r_{1,2}(r_2-r_1)/\nu $.  Asterisks mark Rayleigh-unstable flows;
  squares, quasi-Keplerian flows, \emph{i.e.}  $\partial |\Omega| /\partial r<0$ but
  $\partial |r^2\Omega| /\partial r>0$; diamonds, solid-body flows ($d\Omega/dr=0$);
  crosses for the inner cylinder at rest; triangles for flows explored
  in previous experiments\citep{taylor36,wendt33,schultz59}; and the
  rectangular box for the parameter regime explored by Richard\citep{richard01}.
  Dashed lines denote constant values of $q\equiv - \pd \ln\Omega / \pd \ln r$ 
  at $r=17$~cm, where most of our measurements of
  Reynolds stress were performed (Fig.\ref{figure:beta_z}).
  Rayleigh-unstable flows exhibit $5-10$\% fluctuations that are
  insensitive to the end-ring speeds. Quasi-keplerian and other
  Rayleigh-stable flows are more sensitive.
  For example, when the
  end rings are fixed to the cylinders, fluctuations up to
  $4-8\%$ occur. When the end-ring speeds are adjusted so that
  the ideal-Couette profile is restored,
  fluctuations are $\lesssim 1-2$\% and indistinguishable from those of our 
  solid-body flows, except within a few cm of the boundary.
  Reducing $Re$ by a factor
  $\sim 18$, using an admixture of glycerol, actually \emph{increases}
  the fluctuation level for the same (nearly ideal-Couette) profile.  We
  interpret this to mean that the residual unsteady secondary circulation
  penetrates deeper into the bulk flow at lower $Re$.
  We infer from this that the experiments by Richard\citep{richard01} may have been
  affected by the endcaps, although his ratio of height to gap width
  exceeded ours. His endcaps were split but fixed to the cylinders.}
\label{figure:diagram}
\end{figure}

Experimental flows studied by ourselves and others are summarized in
Fig. \ref{figure:diagram}. Our Reynolds numbers are up to 20
times larger than those previously achieved by Richard\citep{richard01}.  
A Laser Doppler Velocimetry (LDV) system
was employed to sample the azimuthal velocity $v_\theta$
at various radial and axial locations. Both mean values,
$\overline v_\theta\equiv r\Omega$, and fluctuations, $v_\theta^\prime \equiv
v_\theta - \overline v_\theta$, were obtained.
At our Reynolds numbers, linearly-unstable
flows are always turbulent, with fluctuation levels
$(\overline{v_\theta^{\prime\,2}})^{1/2}/\overline{ v}_\theta
=5-10\%$. They are largely insensitive to the endcap speeds.  In
contrast, quasi-Keplerian and other linearly-stable flows are
sensitive to the endcap boundary conditions.  When the endcap speeds
are adjusted to best approximate ideal Couette flow, the fluctuations
are $\lesssim 1-2$\% and indistinguishable from those of our
solid-body flows,  which are expected to be
steady or laminar due to the lack of shear to drive turbulence.

One hallmark of nonlinear instability
is hysteresis: the transition from laminar flow to
turbulence occurs at higher Reynolds
numbers than the reverse\citep{richard01}.
Quiescent flows were gradually brought
into linearly-unstable regimes by raising $\Omega_1$ or lowering
$\Omega_2$, then returned to linearly stable regimes.  Significant
fluctuations were found only in the linearly unstable regime; no
hysteresis was detected.

In the absence of magnetic fields, turbulent angular momentum transport requires
correlated velocity fluctuations.  The radial angular momentum flux is
$\rho r\overline{v_\theta^\prime v_r^\prime}$ where $v_r^\prime$ is the fluctuation in radial velocity, 
and the turbulent viscosity $\nu_\text{turb}$ is defined by equating this 
to $-\rho\nu_\text{turb} r^2\pd\Omega/\pd r$ where
$\nu_\text{turb} = \beta \left| r^3 {\pd \Omega/ \pd r}\right|$, so
that $\beta={\overline{v_\theta^\prime v_r^\prime}}/{ (r^2 \pd \Omega/\pd r)^2}$ is dimensionless.
Conveniently, in a turbulent but statistically steady state with profile (\ref{Couette}),
both $\beta$ and $\nu_\text{ turb}$ are radially constant.
A value of $\beta = (1-2)\times 10^{-5}$ has been deduced \citep{richard99}
from experiments \citep{wendt33,taylor36} with
$0\leq \Omega_1/\Omega_2<1$.

\begin{figure}[]
\centerline{
\includegraphics[width=3.5truein]{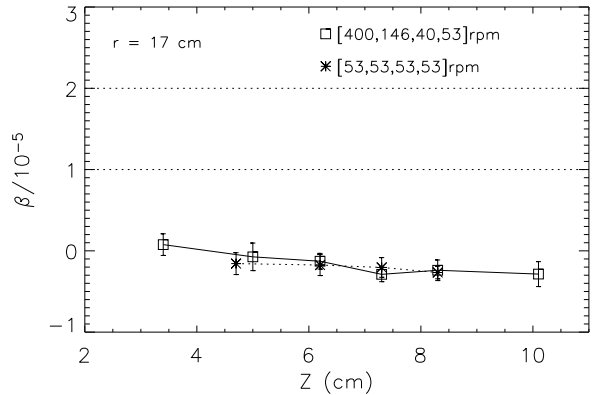}}
\caption{\textbf{Experimentally measured Reynolds stress versus height in
    a quasi-Keplerian profile.}  Here $\beta \equiv
  \overline{v_\theta^\prime v_r^\prime}/ (\overline{v_\theta}^2
  q^2)\,\mbox{sign}(q\Omega)$, where $q\equiv -\partial \ln\Omega/\partial \ln r$:
  $q=+3/2$ in Keplerian
  disks, $1.2\le q\le 1.9$ in our quasi-Keplerian flows.  Square
  symbols connected by solid line were taken at $q=1.86$ (see Fig.\ref{figure:diagram}).  
  Starred points
  connected by dotted line are a solid-body case ($q=0$), which should
  be nonturbulent and therefore serves as a control for systematic errors.  The measured
  values fall far below the range of $\beta$ proposed by Richard \&
  Zahn\citep{richard99}, shown with horizontal dotted lines.  The
  measurements were performed using a synchronized, dual
  laser-Doppler-velocimetry (LDV) system, which allows simultaneous
  detection of both components of $v_\theta$ and $v_r$. The laser
  beams enter the fluid vertically through the acrylic endcaps from
  below (see Fig.\ref{figure:setup}), and $Z$ is the height above
  lower endcap.  Rotation speeds
  $[\Omega_1,\Omega_3,\Omega_4,\Omega_2]$ are shown, where $\Omega_3$
  and $\Omega_4$ are the angular velocities of the inner and outer end
  rings.  Apparently gaussian deviations amounting to $\sim (1-2\%)$ 
  of $\overline{v_\theta}$ were observed
  from sample to sample, representing measurement errors and perhaps true fluctuations.
  Each data point is the mean of $\beta$ computed from $3,000-10,000$ samples.
  Error bars represent the standard deviations.}
\label{figure:beta_z}
\end{figure}

\begin{figure}[]
\centerline{
\includegraphics[width=3.2truein]{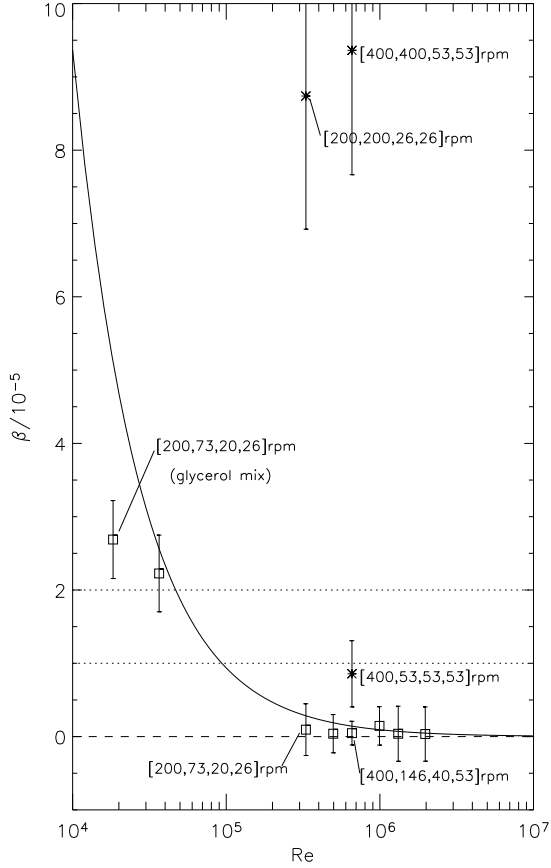}}
\caption{\textbf{Dimensionless Reynolds stress at Reynolds numbers up to $2 \times 10^6$.}
At the largest $Re$, the data show no sign that $\beta [=
\overline{v_\theta^\prime v_r^\prime}/ (\overline{v_\theta}^2 q^2)]$
or the relative fluctuations $\overline{v^{\prime\,2}}/(\overline{v_\theta})^2$
themselves increase with $Re$ in quasi-Keplerian flows.
Here $q\equiv-\partial \ln\Omega/\partial \ln r$ is calculated from the mean $v_\theta (r)$ profiles.
Systematic errors in $\beta$ have been accounted for
by reference to identical measurements in solid-body flows.
Different sizes in error bars (showing s.d.) for the 6 quasi-Keplerian flows with 
optimum boundary conditions (squares at $Re>10^5$) are largely
due to different numbers of measurement samples, $N$.
Averaging over these 6 points, weighted by $\sqrt{N}$, results in
$\beta = 0.72 \times 10^{-6}$ with s.d. of  $2.7\times 10^{-6}$ or
$\beta<6.2 \times 10^{-6}$ at 2 s.d. ($98\%$ confidence).
When the end-ring speeds are not optimized to produce Couette profiles (\ref{Couette}),
however, $\beta$ exhibits large values (asterisks).
When optimal speeds are used but
glycerol is added to the water to reduce the Reynolds number, larger $\beta$ values
are also seen (squares at $Re<10^5$),
possibly due to stronger residual secondary circulation.
For reference, the solid line is the transport expected due to molecular viscosity:
 $\beta_\text{visc} \equiv \nu / [\bar r^3
  (\Omega_2-\Omega_1)/(r_2-r_1)]=Re^{-1} (r_2-r_1)^2/\bar r^2$.
  The proposed $\beta$ lies between the horizontal dotted lines \citep{richard99}. }
\label{figure:beta_re}
\end{figure}

We have measured the Reynolds stress directly
using a synchronized dual-LDV system.  
Both $v_\theta^\prime$ and $v_r^\prime$ appear to follow
Gaussian statistics (E.S. et al., manuscript in preparation),
and random errors were reduced by averaging
$10^3$ to $10^4$ samples.  Systematic errors were gauged by
comparison with solid body flows
($\Omega_1=\Omega_2=\Omega_3=\Omega_4$).  Figure
\ref{figure:beta_z} shows that $\beta$ differs indistinguishably
between quasi-Keplerian and solid-body rotation, falling far below the 
proposed range \citep{richard99}.
A large outward Reynolds stress 
is detected in linearly-unstable flows, $\beta >10^{-3}$,
as expected 
(M.B. et al., manuscript in preparation). Even for linearly-stable flows,
when the speeds $(\Omega_3,\Omega_4)$ of the endcaps were not adjusted properly 
to produce the ideal-Couette profile (\ref{Couette}),  
$\beta $ was found to be almost $10^{-4}$ (Fig.~\ref{figure:beta_re}).
This again indicates that the axial boundaries can profoundly influence
linearly-stable flows, as previously suggested\citep{schultz59,lesur_longaretti05}.

A final point should be made about the critical Reynolds number for
transition, $Re_\text{crit}$, versus gap width.  
Based on the experiments of Wendt and Taylor \citep{wendt33,taylor36},
the scaling $Re_\text{crit} \simeq 6\times 10^5 \cdot [(r_2-r_1)/\overline r]^2$ 
has been proposed\citep{zeldovich81,dubrulle93,richard99,longaretti02}.
It is unclear whether this scaling applies to quasi-Keplerian flow since
it was derived from data for $\Omega_2/\Omega_1>1$.
In any case, up to $Re=2\times 10^6$, which is about
three times the proposed $Re_\text{crit}$ since $(r_2-r_1)/\overline r \sim 1$ in our
device, we have seen no signs of rising fluctuation levels or
Reynolds stress (Fig.~\ref{figure:beta_re}).

Therefore, we have shown that purely hydrodynamic quasi-Keplerian
flows, under proper boundary conditions {\it and} at large enough
Reynolds numbers, cannot transport angular momentum at astrophysically
relevant rates.

Of course, it could be argued that our maximum $Re$,
which only barely exceeds some theoretical estimates of
$Re_\text{crit}$ \citep{mukhopadhyay05}, is still not large enough for
transition.  Or it could be that transition has occurred, but that the
transport is too small for us to detect.
To extrapolate from $Re\le 2\times 10^6$ to a typical astrophysical
value $\gtrsim 10^{12}$, we rely on the empirical observation that for
$Re>Re_\text{crit}$, the Reynolds number based on the \emph{turbulent}
rather than the molecular viscosity is approximately independent of $Re$
itself \citep{colebrook39}.  It follows that
$\nu_\text{turb}\approx LU/Re_\text{crit}$ for $Re>Re_\text{crit}$, where $L$ and $U$ are the
characteristic size and velocity of the flow.  It is
common knowledge among civil engineers that this is true of flow in pipes, for example.
If it is true of
rotating shear flow, as theoretical arguments suggest it should be
\citep{lesur_longaretti05}, then $\beta$ at $Re\gtrsim 10^{12}$ should
be comparable to what we find at $Re \lesssim 2\times 10^6$, namely
%$\beta = 0.72 \times 10^{-6}$ with s.d. of  $2.7\times 10^{-6}$ or
$\beta<6.2 \times 10^{-6}$ (at 2 s.d., or $98\%$ confidence; see Fig.\ref{figure:beta_re} caption)
---whether or not we have crossed the threshold of transition.

Lastly, it is useful to relate the above upper bound for $\beta$ to
the more commonly used Shakura-Sunyaev $\alpha$
parameter\citep{shakura73}: $\nu_\text{turb}= \alpha \Omega h^2$.  We replace
this by $\nu_\text{turb}=\alpha \Omega (r_2-r_1)^2$ since $r_2-r_1$ is smaller
than $h$ in our experiment, and we presume that the dominant turbulent eddies 
scale with the smallest dimension of the flow ($h\ll r$ in most disks).  Then
$\alpha =\beta q \overline r^2/(r_2-r_1)^2 \simeq \beta q$ (see Fig.\ref{figure:beta_z} caption for
a definition of $q$), and thus
our results imply a similar upper bound for $\alpha$ as for $\beta$ in
purely hydrodynamic disks, whereas protostellar-disk lifetimes and accretion
rates indicate\citep{hartmann98} $\alpha\gtrsim 10^{-3}$.  Although some have suggested that
complications such as vertical or radial stratification may yet lead
to essentially \emph{linear} nonaxisymmetric hydrodynamic
instabilities\citep{klahr03,dubrulle05}, our belief is
that such nonaxisymmetric linear instabilities depend upon radial
boundaries and hence are not generally important in thin
disks\citep{johnson06,goodman_balbus01}.  If this is correct, then by
default, the magnetorotational instability appears to be the only plausible source of accretion disk
turbulence.

\vspace{1cm}
\noindent
{\bf Received 8 August; accepted 6 October 2006}

\bibliographystyle{unsrtnat}

\noindent
{\bf Acknowledgments} We thank S. Balbus for discussions, R. Cutler for technical assistance on the apparatus, P. Heitzenroeder, C. Jun, L. Morris, and S. Raftopolous for engineering assistance, as well as Dantec Dynamics for the contracted use of a LDV measurement system. This research was supported by the US Department of Energy, Office of Science Ð Fusion Energy Sciences Program; the US National Aeronautics and Space Administration, Astronomy and Physics Research and Analysis and Astrophysics Theory Programs; and the US National Science Foundation, Physics and Astronomical Sciences Divisions.

\noindent
{\bf Author Contributions}
H.J., M.B., and E.S. planned and executed the experiments, and analyzed data; E.S. and M.B. prepared apparatus and diagnostics; H.J. drafted the paper; J.G. suggested this subject and assisted in the interpretation of its results and in revising the paper. 

\noindent
{\bf Author Information} Reprints and permissions information is available 
at npg.nature.com/reprintsandpermissions. The authors declare no competing
financial interests. Correspondence and requests for materials
should be addressed to H.J. (hji@pppl.gov).

\end{document}